\documentclass[aps,10pt,groupedaddress,twocolumn,superscriptaddress,nofootinbib,pre]{revtex4-1}

\usepackage{graphicx}
\usepackage{color}
\usepackage{amsmath}
\usepackage{amsfonts}
\usepackage{amssymb}
\usepackage{dcolumn}
\usepackage{booktabs}
\usepackage{paralist}
\usepackage[acronym]{glossaries}

\newacronym{WAN}{WAN}{World Airline Network}
\newacronym{MST}{MST}{Minimum Spanning Tree}

\begin{document}

\title{Model for the growth of the World Airline Network}

\author{T. Verma} 
\affiliation{ ETH
  Z\"urich, Computational Physics for Engineering Materials, Institute
  for Building Materials, Wolfgang-Pauli-Strasse 27, HIT, CH-8093 Z\"urich,
  Switzerland}

\author{N. A. M. Ara\'ujo}\email{nmaraujo@fc.ul.pt} \affiliation{Departamento de F\'isica, Faculdade de Ci\^encias, Universidade de Lisboa, P-1749-016 Lisboa, Portugal, and Centro de F\'isica Te\'orica e Computacional, Universidade de Lisboa, 1749-016 Lisboa, Portugal}

\author{J. Nagler} \affiliation{ ETH
  Z\"urich, Computational Physics for Engineering Materials, Institute
  for Building Materials, Wolfgang-Pauli-Strasse 27, HIT, CH-8093 Z\"urich,
  Switzerland}
  
\author{J. S. Andrade Jr.} \affiliation{ ETH
  Z\"urich, Computational Physics for Engineering Materials, Institute
  for Building Materials, Wolfgang-Pauli-Strasse 27, HIT, CH-8093 Z\"urich,
  Switzerland} \affiliation{Departamento de F\'isica, Universidade
  Federal do Cear\'a, Campus do Pici, 60455-760 Fortaleza, Cear\'a,
  Brazil}

\author{H. J. Herrmann} \affiliation{ ETH
  Z\"urich, Computational Physics for Engineering Materials, Institute
  for Building Materials, Wolfgang-Pauli-Strasse 27, HIT, CH-8093 Z\"urich,
  Switzerland} \affiliation{Departamento de F\'isica, Universidade
  Federal do Cear\'a, Campus do Pici, 60455-760 Fortaleza, Cear\'a,
  Brazil}


\begin{abstract}
We propose a probabilistic growth model for transport networks which employs a balance between popularity of nodes and the physical distance between nodes. By comparing the degree of each node in the model network and the \gls{WAN}, we observe that the difference between the two is minimized for $\alpha\approx 2$. Interestingly, this is the value obtained for the node-node correlation function in the WAN. This suggests that our model explains quite well the growth of airline networks.
\end{abstract}

\maketitle

\section{Introduction}
The real world runs with overhead costs. In recent times, with the advent of technology and infrastructure systems, most transport networks need not be constrained anymore by the lack of global routing information. Instead a defined cost must direct its development. The \gls{WAN} is a superimposition of local networks established at different points in time in different locations around the globe. It has evolved through political ties between countries and trade relations across borders while simultaneously following the development of cities and using global routing information \cite{Bairoch1985}. 

Navigation in the real world depends on this global knowledge. The World Airline Network is a core-periphery network \cite{Verma2014} with long-range connections that make possible its relatively small diameter compared to its size. The existence of central hubs that form the interconnected core, together with the long-range connections that separate the hubs to span across the globe, gives the \gls{WAN} its unique topology \cite{Csermely2013a,Barrat,Grubesic2008,Rossa,Verma2014,Wilkinson2011}. 

So far, researchers have analyzed the structure of the \gls{WAN} and other transport networks, discussed ways to improve it, and proposed a simple model to understand the plausible mechanism behind its core-periphery structure \cite{Louf,Louzada2015,Peixoto2012a,Verma2014,Verma2016}. Here, we introduce a probabilistic law: addition of long-range connections to a network lattice constructed from the real world positions of all the airports in the \gls{WAN} considering its Euclidean distances, passenger flows and flight connections. This growth model allows us to study the features of a network that is embedded in space and has a cost of construction and maintenance attached to it. We demonstrate the existence of a family of networks that resemble closely a real-world scenario by optimizing certain navigability conditions. 

The paper is organized in the following way. In the next section we will present the model for the growth of the \gls{WAN}. Following this, we will show results for our model with respect to changes in the tuning parameter $\alpha$. The final section will discuss the findings and show how our model can recover parts of the \gls{WAN}. 

\section{Model}
Consider the world airline network. We know the location of its nodes (latitudinal and longitudinal coordinates of the airports \cite{OpenFlights}) and the Euclidean distance between them, namely $d_{ij}$. 

\begin{figure}
\includegraphics[width=0.5\textwidth]{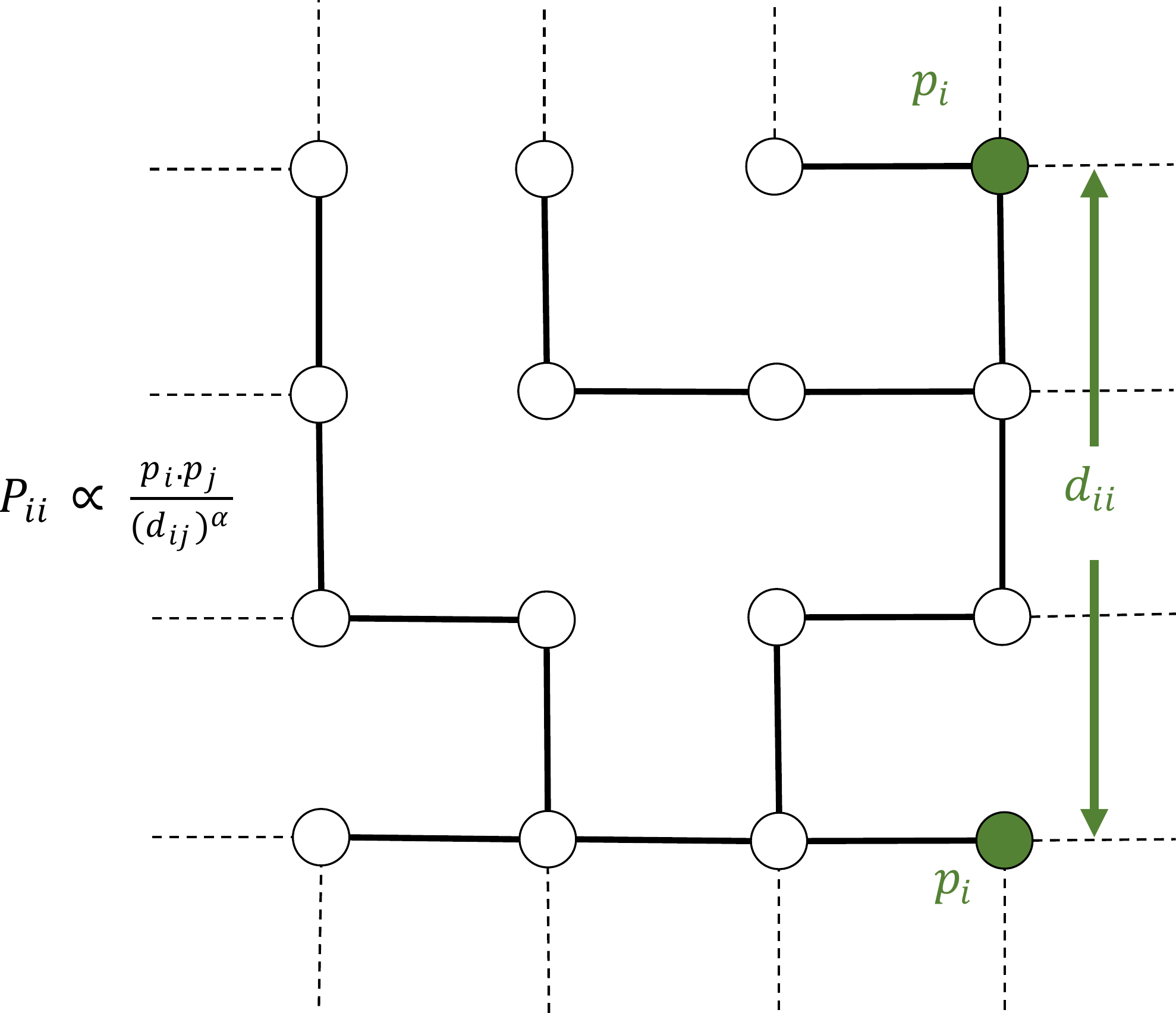}
\caption[Growth Model for the addition of long-range connections]{{\bf Growth Model showcasing the addition of long-range connections.} Nodes $i$ and $j$ are assigned popularities $p_i$ and $p_j$. The distance between nodes is the Euclidean distance $d_{ij}$ between them. $i$ and $j$ are randomly selected for long-range connections based on the probabilistic law $P_{ij}$.}
\label{fig:GrowthModel}
\end{figure}

The \gls{WAN} is comprised of $N = 3237$ nodes representing airports and $L = 18125$ links depicting direct connections between the airports. A growth model of this sort can be initialized as a weighted \gls{MST} over the \gls{WAN} with only $N - 1 = 3236$ links with weights being the Euclidean distances between them. We start with a spanning tree to guarantee that we have only one large component that includes all nodes. The \gls{MST} establishes that our starting network is the most efficient in terms of costs and/or traveling time. Note that there is a unique \gls{MST} for networks with different weights on every link \cite{Sedgewick2011}. Hence, every node $i$ is connected with every other node through the shortest path in the network. The model is illustrated in fig. \ref{fig:GrowthModel}.

We begin by introducing our growth model as a means to add long-range connections to the \gls{MST}, in a similar manner as previously introduced on navigation through small-world spatial networks~\cite{Roberson2006,Rozenfeld2010,Li2010,Barthelemy2011,Li2013,Oliveira2014}. Pairs of nodes $ij$ are randomly selected for adding long-range connections. These connections are chosen with a probability $P_{ij}$,
\begin{equation}
P_{ij} \propto \frac{p_i.p_j}{{d_{ij}}^\alpha},
\end{equation}
where $p_i$ is the popularity of node $i$ (details below) and $\alpha$ is a tuning parameter, sometimes called the clustering exponent \cite{Kleinberg2000}. Probability of adding a link then decays with the distance between the nodes as expected for transport networks.

The popularity of a node is related to the relevance of the node in the network. Since we want to compare to the \gls{WAN}, we considered the popularity to be equal to the number of passengers serviced by the corresponding airport in the year $2011$ \cite{Verma2014}. 

The process of adding long-range connections is constrained by a budget function $B$. The budget function,
\begin{equation}
B = \sum_{i,j \mbox{ } i\neq j} {d_{ij}},
\end{equation} 
restricts the growth of the network beyond the underlying overall link length of the \gls{WAN}. This is a measure to keep the size of the model network close to the real scenario as a real infrastructure network cannot construct infinitely many links in order to optimize its navigability. 

The question that we address here is: for which value of $\alpha$ the degree of the nodes is closest to the one in the \gls{WAN}?

\section{Results}
We studied the relationship between our model networks and the \gls{WAN} using bootstrapping techniques \cite{Bland1996,Varian}. To compare to the \gls{WAN} we measured the weighted standard deviation of bootstrapped samples, which is calculated as follows,
\begin{equation}
\frac{ \sum_{i=1}^N w_i (x_i - \bar{x}^*)^2 }{N},
\end{equation}
where $N$ is the number of observations, $w_i$ are the weights, $x_i$ are the observations and $\bar{x}^*$ is the weighted mean. In our case, $N = 10^4$, $x_i$ is the model measurement, $\bar{x}^*$ is the \gls{WAN} measurement, and $w_i = \frac{1}{rank(WAN)}$ where the rank is calculated using the degree of the nodes: the highest degree node is ranked $1$, the second highest as $2$ and so on and so forth.

\begin{figure*}
\centering
\includegraphics[width=0.45\textwidth]{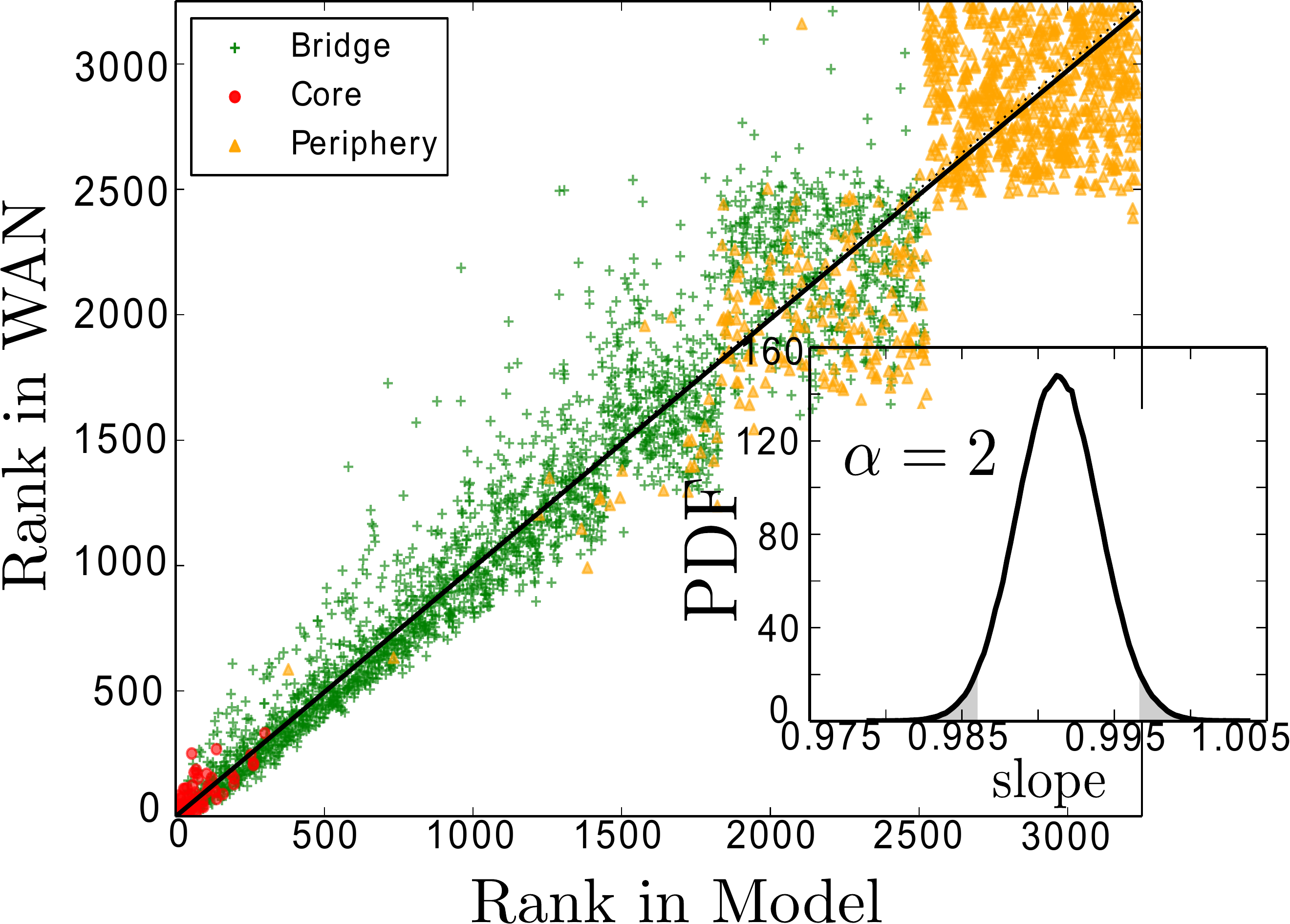}
\includegraphics[width=0.4\textwidth]{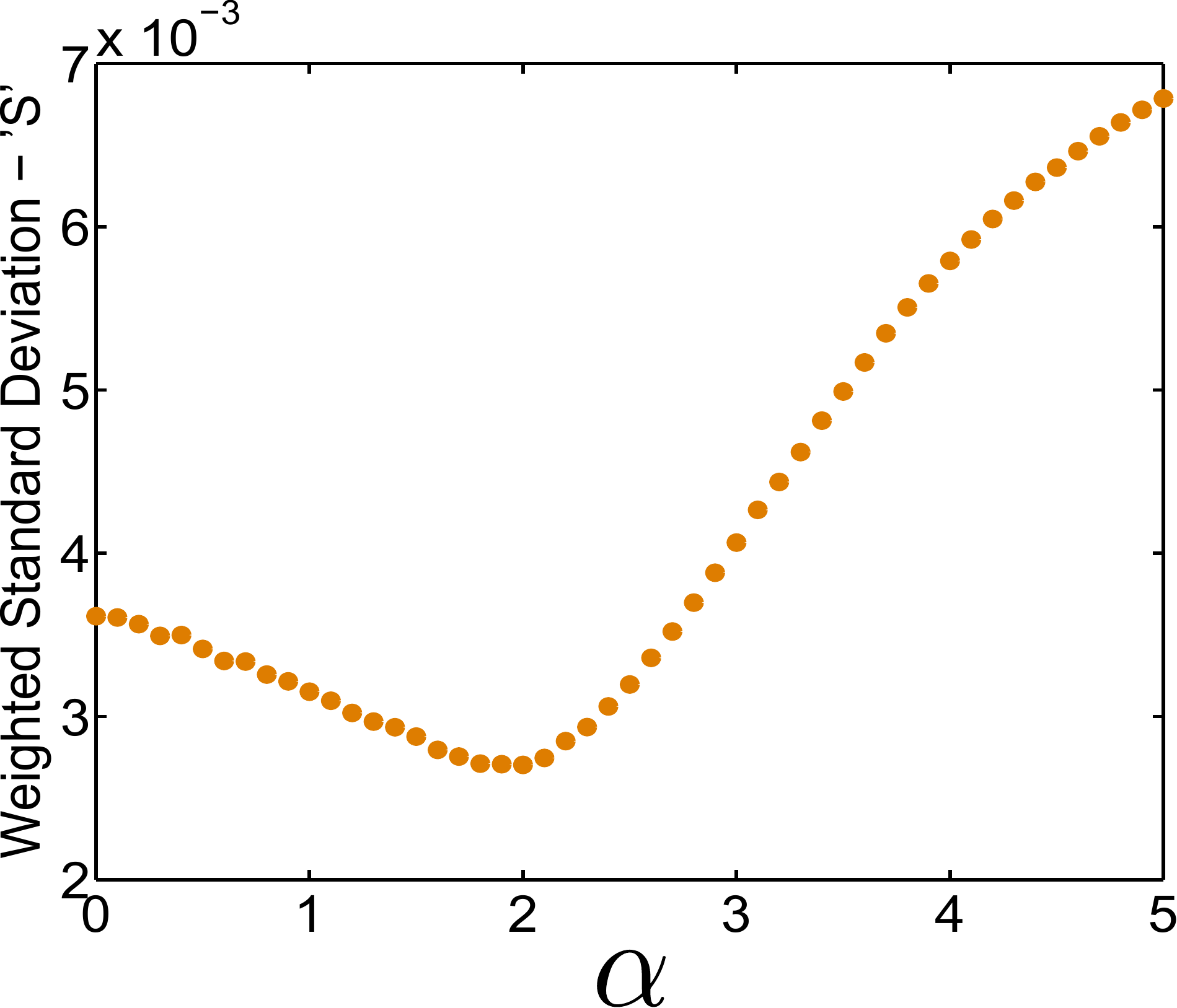}
\caption[Bootstrapping analysis of a random sampling of networks with the \gls{WAN}]{{\bf Bootstrapping analysis of a random sampling of networks with the \gls{WAN} for $\alpha \approx 2$.} Figure a) shows the correlation in rank of nodes per degree between the model and the \gls{WAN}. The inset shows a highly accurate match of data sampling. Figure b) illustrates the relation between the weighted standard deviation of the samples with respect to changes in $\alpha$. Clearly, there is a minimum at $\alpha \approx 2$. Sampling data are based on $10^4$ realizations.}
\label{fig:bootstrap}
\end{figure*}

In Fig. \ref{fig:bootstrap}a, for $\alpha \approx 2$ there is a minimum in the weighted standard deviation, suggesting that, it is for this value of $\alpha$ that node degrees are closest to the ones in the \gls{WAN}. The color codes are depicting core (red, circle), periphery (yellow, triangle) and the remaining bridge (green, cross) of the network \cite{Verma2014}. Figure \ref{fig:bootstrap}b shows the relationship between weighted standard deviation \cite{Bland1996} of the samples with $\alpha$. The standard deviation is inversely dependent on the rank of the nodes and hence nodes in the core contribute much more than the peripheral nodes to this measurement. 

\begin{figure*}
\centering
\includegraphics[width=0.75\textwidth]{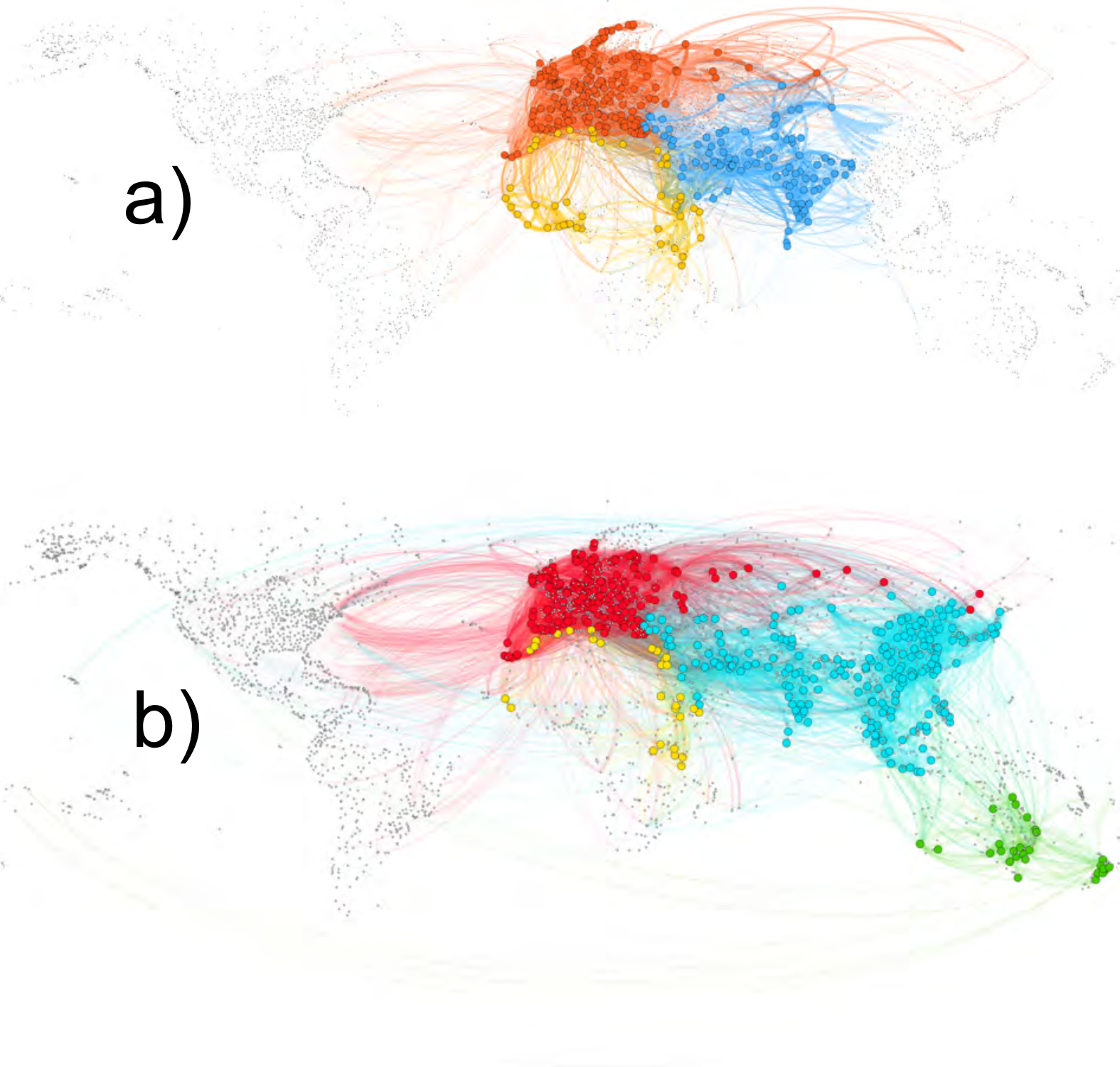}
\caption[Visualization of a network obtained with the growth model]{{\bf Visualization of a replicated network from the growth model.} Figure a) shows the network of nodes (and their links) that form the largest connected component through links that exist in $100\%$ of all $100$ data samples. Figure b) illustrates the network of nodes (and their links) that form the largest connected component through links that exist in $50\%$ of all $100$ data samples. Naturally, this network spans a larger surface of the globe. Sampling data are based on $100$ realizations.}
\label{fig:replication}
\end{figure*}

Figure \ref{fig:replication}a shows the network of nodes (and their links) that form the largest connected component ($\approx 20\%$ of the \gls{WAN}) through links that exist in all the $100$ data samples that we generated with our model. Figure \ref{fig:replication}b illustrates the network of nodes (and their links) that form the largest connected component ($\approx 32\%$ of the \gls{WAN}) through links that exist in $50\%$ of all $100$ data samples.

\section{Discussion}
The presence of a minimum in the weighted standard deviation reiterates that our choice of probabilistic law for the growth model with $\alpha = 2$ allows us to create networks that resemble the \gls{WAN} to a large extent. The distance between airports, $d^{-\alpha}$, decays in the optimal case as a power law for $\alpha \approx 2$ \cite{Louzada2015}, which could be understood in terms of of Kleinberg's theorem \cite{Kleinberg2000}.

Both considerations, budget restrictions and geographical conditions contribute to the development of the \gls{WAN}. Our efforts to add a budget restriction bring a sense of reality to the modeling approach while the network is simultaneously restricted by its geography; which explains the use of real world distances for navigation. Our results show that the tuning parameter is the key factor for selection of long-range connections. Increasing this value favors shorter distances and thus heavier loads as a consequence of the popularity of the selected nodes. This accomplishes easy routing and results in more clustering but the small world nature of the network breaks down. At the critical tuning parameter, the small world nature of our model networks remains intact because of the decaying of distances between airports. 

Our approach might be used for different transport networks that may exhibit different decaying laws. 

\begin{acknowledgments}
We acknowledge financial support from the ETH Risk Center with grant ETH48, European
Research Council through Grant FlowCSS No. FP7-319968. and Portuguese Foundation for Science and Technology (FCT) under Contracts nos. EXCL/FIS-NAN/0083/2012, UID/FIS/00618/2013 and IF/00255/2013.

\end{acknowledgments}

\bibliography{Thesis}

\end{document}